\documentclass[aps,prd]{revtex4}
\usepackage{amsfonts}
\usepackage{amsmath}
\usepackage{amssymb}
\usepackage{graphicx}

\input{tcilatex}
\begin{document}

\title{ GRAVITATIONAL LENSING EFFECT ON THE HAWKING RADIATION OF DYONIC
BLACK HOLES}
\author{I.\ Sakalli}
\email{izzet.sakalli@emu.edu.tr}
\author{A. Ovgun}
\email{ali.ovgun@emu.edu.tr}
\author{S.F. Mirekhtiary}
\email{fatemeh.mirekhtiary@emu.edu.tr}
\affiliation{Physics Department, Eastern Mediterranean University, G.Magosa North Cyprus,
Mersin-10, Turkey}
\date{01.02.2014}

\begin{abstract}
In this paper, we analyze the Hawking radiation (HR) of a non-asymptotically
flat (NAF) dyonic black hole (dBH) in four-dimensional (4D)
Einstein-Maxwell-Dilaton (EMD) gravity by using one of the semiclassical
approaches which is the so-called Hamilton-Jacobi (HJ)\ method. We
particularly motivate on the isotropic coordinate system (ICS) of the dBH in
order to highlight the ambiguity to be appeared in the derivation of the
Hawking temperature ($T_{H}$) via the HJ method. Besides, it will be shown
that the ICS allows us to write the metric of the dBH in form of the Fermat
metric, which renders possible of identification of the refractive index ($n$%
) of the dBH. It is unraveled that the value of $n$ and therefore the
gravitational lensing effect is decisive on the the tunneling rate of the
HR. We also uncloak how one can resolve the discrepancy about the $T_{H}$ of
the dBH in spite of that lensing effect.
\end{abstract}

\maketitle

\section{Introduction}

In classical point of view, black holes (BHs) are such objects that due to
their huge attractive forces even the light cannot escape from them. But, if
we take the quantum effects into account, they could be gray i.e., not
entirely black. That surprising claim was made by Hawking \cite%
{Hawking,Hawking2} over forty years ago. Hawking thought that the quantum
mechanics should allow the particles to tunnel through the event horizon of
the BH. After making marvelous calculations, Hawking theoretically showed
that BHs should emit a steady flux of thermal radiation with a temperature $%
T_{H}=\frac{\kappa }{2\pi }$ in which $\kappa $ is the gravitational field
strength at the event horizon. On this basis, a number of new methods have
been proposed (a reader may consult \cite{DonPage,Vanzo} for the topical
review). \ 

Among those methods, the universal consent one is the tunneling method which
is devised by Kraus and Wilczek (KW) \cite{KW1,KW2}. KW proved that the HR
can be thought as a dynamical model in which while BH shrinks, the particles
radiate. In this dynamical model, both energy conservation and
self-gravitational effects which were not considered in the original
derivation of HR \cite{Hawking,Hawking2} are taken into account. In fact,
they treated the each radiating particle as a self-gravitating thin
spherical shell with energy $\omega $. Hence, by using the null geodesics of
the associated particles, it became possible to get the action ($I$) (and
henceforth its imaginary part, ${Im}I$) of the tunneling particle, which
yields the tunneling rate as $\Gamma \approx e^{-2{Im}I}$ and admits the $%
T_{H}$ due the well-known quantum mechanical result, $\Gamma \approx
e^{-\omega /T}$. Later on, KW's method was improved by Parikh and Wilczek
(PW) \cite{PW} in order to show that the spectrum of the HR is not pure
thermal, which implies unitarity of the underlying quantum process, and the
resolution of the information loss paradox \cite{ILP1,ILP2}.

Among the other methods, the HJ method which uses the relativistic HJ
equation attracts great interest. This method firstly devised by Angheben \
et al. \cite{Angheben} (and references therein). They indeed built up an
alternative method for calculating ${Im}I$. The HJ method within the WKB
approximation ignores the self-gravitational effect and energy conservation.
Generally, the HJ method is employed by substituting a suitable ansatz into
the relativistic HJ equation. For the separability, the chosen ansatz should
involve the symmetries of the spacetime. The final radial equation is solved
by an integration which is along the classically forbidden trajectory
starting from inside of the BH and extends to the observation point. But
during this calculation, the integral always possesses a pole located at the
horizon. This pole can be circumvented by applying the method of complex
path analysis \cite{Padnab1,Padnab2,Padnab3}.

According to us, most of the BHs in the macrocosm should be portrayed with
NAF geometries. The mainstay of our thoughts is the
Friedmann-Robertson-Walker \cite{FRW}, which is assumed to be one of the
best theoretical models in describing our universe and, as is known, its
geometry is NAF. For this reason, we focus on the commonly acceptable NAF
BHs in order to compute their $T_{H}$ by employing the HJ method. In the
same line of thought, we consider the dBHs which are originally found by
Yazadjiev \cite{YazadCQG}. The dBH solutions considered here have two
horizons hiding a curvature singularity at the origin. As mentioned in \cite%
{YazadCQG}, they may serve as backgrounds for non-supersymmetric holography
and lead to possible extensions of AdS/CFT correspondence \cite{ClementLey}.
On the other hand, when the standard HJ method is applied to the dBH which
can be expressed in the ICS, we encounter with a discrepancy between its
computed horizon's temperature and its standard $T_{H}$. This discrepancy
problem arising during the use of the ICS has become popular anew. Recently,
Schwarzschild\ \cite{GRG12} and the linear dilaton BHs (from now on we
abbreviate it as LDBHs) \cite{SM13} within the ICS have been thoroughly
studied. Inspiring from \cite{SM13}, we shall also calculate the $n$ of the
medium of the dBH. Then, we highlight the effect of $n$ on the tunneling
rate, and consequently on the $T_{H}$. The latter remark could be associated
with the gravitational lensing effect on the HR as in \cite{Belgiorno}.

The structure of the paper is as follows. In Sec. II, we introduce and
review the dBH. We give its some geometrical and thermodynamical features.
For the sake of example, we follow the systematic HJ method for deriving the 
$T_{H}$ of the dBH in its naive coordinates. In Sec. III, we firstly
transform the naive coordinates of the dBH into the ICS. Then, the effect of
the $n$ on the calculation of the ${Im}I$ is explicitly represented. The
obtained horizon temperature is the half of the $T_{H}$. At the last stage
of this section, a detailed analytical analysis is performed in order to
resolve the differences in the temperatures. We draw our conclusions in Sec.
IV.

The paper uses the signature $(-,+,+,+)$ and units where $c=G=\hbar =k_{B}=1$

\section{dBH geometry and HJ Method}

In this section, we shall first describe the dBH spacetime which is the
solution to the 4D EMD gravity. Its line-element represents a static and
spherically symmetric solution in the low-energy limit of the string theory
in which gravity is coupled to the electromagnetic field and dilaton. These
solutions posses both electric and magnetic charges, and they admit NAF
geometry. Then, we will apply the HJ method to the dBH metric given in its
naive coordinates in order to show that it concludes with the $T_{H}$.

In the EMD\ theory \cite{Chan}, the 4D action is given by

\begin{equation}
S=\int d^{4}x\sqrt{-g}(\Re -2g^{\mu \nu }\nabla _{\mu }\varphi \nabla _{\nu
}\varphi -e^{-2\alpha \varphi }F_{\mu \nu }F^{\mu \nu }),  \label{1}
\end{equation}

where $\Re $ denotes the scalar curvature with respect to the spacetime
metric $g_{\mu \nu }$, $\varphi $ is the scalar dilaton field with a
coupling constant $\alpha $ and $F_{\mu \nu }$ is the electromagnetic field.
The dBH solution of the above action is given by \cite{YazadCQG}. Here, we
shall take $\alpha =1$ as it has been considered very recently in \cite%
{Slavov}. The dBH spacetime is described by the following line-element

\begin{equation}
ds^{2}=-fdt^{2}+f^{-1}dr^{2}+R^{2}(d\theta ^{2}+\sin ^{2}\theta d\varphi
^{2}),  \label{2}
\end{equation}

in which the metric functions are 
\begin{eqnarray}
f &=&\frac{(r-r_{+})(r-r_{-})}{r_{0}r},  \notag \\
R^{2} &=&r_{0}r.  \label{3}
\end{eqnarray}

The constants $r_{+}$ and $r_{-}$ represent inner and outer horizons,
respectively. While $r_{+}$ and $r_{-}$ determine the value of the magnetic
charge $Q_{m}$, another constant $r_{0}$ governs the electric charge $Q_{e}$%
. Those relationships are given by

\begin{equation}
r_{0}=\sqrt{2}Q_{e},\text{ \ \ \ \ }r_{+}r_{-}=2Q_{m}^{2},  \label{4}
\end{equation}

When the $Q_{m}$ is vanished by setting the inner horizon to zero i.e., $%
r_{-}=0$, the metric functions (3) describe the pure electrical LDBH
spacetime. The eponyms of the LDBH are Cl\'{e}ment and Gal'tsov \cite%
{Clement}. One of the most interesting features of the LDBHs is that while
performing the HR, they undergo an isothermal process. Namely, their HR is
such a special process that the energy (mass) transfer out of them typically
happens at such a slow rate that thermal equilibrium is always maintained.
Today, there are numerous studies on the LDBHs (see for instance \cite%
{SM13,Pasaoglu,Sakalli1,Sakalli2,MSH,Sakalli3,Sakalli4,RLi}).

Due to the NAF structure of the dBH, one should follow Brown and York's
quasi-local mass ($M$) formalism \cite{BrownYork}. Thus, one derives a
relationship between the horizons and the $M$ as follows%
\begin{equation}
M=\frac{1}{4}(r_{+}+r_{-}),  \label{5}
\end{equation}

The $T_{H}$ is expressed in terms of the surface gravity $\kappa $ \cite%
{Wald} as follows

\begin{equation}
T_{H}=\frac{\kappa }{2\pi }=\left. \frac{\partial _{r}f}{4\pi }\right\vert
_{r=r_{+}},  \label{6}
\end{equation}

After substituting the metric function $f$ (3) into the above equation, $%
T_{H}$ of the dBH yields

\begin{equation}
T_{H}=\frac{r_{+}-r_{-}}{4\pi r_{0}r_{+}},  \label{7}
\end{equation}

In the extremal BH\ solution i.e., $r_{+}=r_{-}$, it can be easily seen that
we obtain zero temperature, which is a well-known issue.

Here, we focus on the problem of a scalar particle which crosses the event
horizon from inside to outside. While it acts this classically forbidden
motion, we ignore the back-reaction and self-gravitational effects. Within
the semi-classical framework, the action $I$ of the particle in question
should satisfy the relativistic HJ equation\ which is given by

\begin{equation}
g^{\mu \nu }\partial _{\mu }I\partial _{\nu }I+m^{2}=0,  \label{8}
\end{equation}

where $m$ is the mass of the scalar particle, and $g^{\mu \nu }$ represents
contravariant form of the metric tensors of Eq. (2). By considering Eqs.(2),
(3) and (8), we get%
\begin{equation}
\frac{-1}{f}(\partial _{t}I)^{2}+f(\partial _{r}I)^{2}+\frac{1}{R^{2}}\left[
(\partial _{\theta }I)^{2}+\frac{1}{\sin ^{2}\theta }(\partial _{\varphi
}I)^{2}\right] +m^{2}=0,  \label{9}
\end{equation}

The technique of separation of variables is the commonly used method for the
HJ\ equation. To this end, the $I$ is chosen as follows

\begin{equation}
I=-Et+W(r)+J(x^{k}),  \label{10}
\end{equation}

So, one finds

\begin{equation}
\partial _{t}I=-E,\text{ \ \ \ \ \ }\partial _{r}I=\partial _{r}W(r),\text{
\ \ \ \ \ }\partial _{k}I=J_{k},  \label{11}
\end{equation}

where $J_{k}$'s are constants in which $k=1,2$ label angular coordinates $%
\theta $ and $\varphi $, respectively. The norm of the timelike Killing
vector $\partial _{t}$ becomes (negative) unity at a particular location:

\begin{equation}
r\equiv R_{0}=\frac{1}{2}\left[ r_{+}+r_{-}+r_{0}\left( 1+\sqrt{\frac{%
(r_{+}-r_{-})^{2}}{r_{0}^{2}}+2\frac{(r_{+}+r_{-})}{r_{0}}+1}\right) \right]
,  \label{12}
\end{equation}

It means that when a detector of an observer is located at $R_{0}$ which is
outside the horizon, the energy of the particle measured by the observer
corresponds to $E$. After substituting Eq. (11) into Eq. (9) and solving it
for $W(r)$, we find out that

\begin{equation}
W(r)=\pm \int {\frac{\sqrt{E^{2}-\frac{f}{R^{2}}[J_{\theta }^{2}+\frac{%
J_{\varphi }^{2}}{sin^{2}(\theta )}+\left( mR\right) ^{2}]}}{f}}dr,
\label{13}
\end{equation}

The positive and negative signs appeared in the above equation is due to the
quadratic form of Eq. (9). Solution of Eq. (13) with "$+$" signature
corresponds to outgoing scalar particles and the other solution i.e., the
solution with "$-$" signature refers to the ingoing particles. Evaluating
the above integral around the pole at the horizon (following the Feyman's
prescription given by \cite{Padnab1,Padnab2,Padnab3,Feynman}), one reaches to

\begin{equation}
W(r)\cong W_{\left( \pm \right) }=\pm W_{h}+\chi ,
\end{equation}

where $W_{h}=\frac{i\pi Er_{+}r_{0}}{r_{+}-r_{-}}$ and $\chi $ is a complex
integration constant. Thus, we can deduce that ${Im}I$ arises from the pole
at the horizon and the complex constant $\chi $. Thence, we can determine
the probabilities of ingoing and outgoing particles while crossing $r_{+}$ as

\begin{equation}
P_{out}=e^{-2{Im}I}=\exp \left[ -2{Im}W_{\left( +\right) }\right] ,
\label{15}
\end{equation}

\begin{equation}
P_{in}=e^{-2{Im}I}=\exp \left[ -2{Im}W_{\left( -\right) }\right] ,
\label{16}
\end{equation}

In the classical point of view, a BH absorbs any ingoing particles passing
its horizon. In other words, there is no reflection for the ingoing waves
which corresponds to $P_{in}=1$. This is enabled by setting $W_{\left(
-\right) }=0$ which amounts to ${Im}\chi =\frac{\pi Er_{+}r_{0}}{r_{+}-r_{-}}
$. This choice signifies also that the ${Im}I$ for a tunneling particle
solely depends on $W_{(+)}$. Namely, we get 
\begin{equation}
{Im}I={Im}W_{(+)}={2\pi Er_{+}r_{0}}{r_{+}-r_{-}},  \label{17}
\end{equation}

Therefore, the tunneling rate for the dBH can be obtained as

\begin{equation}
\Gamma =P_{out}=\exp \left( -\frac{4\pi Er_{+}r_{0}}{r_{+}-r_{-}}\right) ,
\label{18}
\end{equation}

and according to \cite{PW}

\begin{equation}
\Gamma =e^{-\beta E},  \label{19}
\end{equation}

in which $\beta $ denotes the Boltzmann factor and $T=\frac{1}{\beta }$, one
can easily read the horizon temperature of the dBH as 
\begin{equation}
\check{T}_{H}=\frac{r_{+}-r_{-}}{4\pi r_{+}r_{0}}.  \label{20}
\end{equation}

This is equal to the $T_{H}$ obtained in Eq. (7).

\section{Isotropic Coordinates of the dBH and Lensing Effect on its HR}

The universe is homogeneous and isotropic on large scales. Isotropy of the
universe is well known from experiments such as distribution of radio
galaxies at Very Large Array in Mexico, Cosmic X-Ray Background and Cosmic
Microwave Background at Planck and WMAP satellites \cite{BookUniverse}. In
this section, we are deal with the ICS. Isotropic coordinates are used in
general to kill the horizon and make the time direction a Killing vector.
Furthermore, time slices appear as Euclidean with a conformal factor. By
using this transformation, the index of refraction, $n,$ of light rays
around the BH\ can be calculated.

The dBH solution in isotropic coordinates \cite{Eddington} can be found by
the following transformation

\begin{equation}
r=r_{-}+\frac{1}{4\rho }\left( \rho +\rho _{h}\right) ^{2}  \label{21}
\end{equation}

where $\rho _{h}=r_{+}-r_{-}$ defines the horizon in the ICS. By making this
transformation, the metric (2) becomes

\begin{equation}
ds^{2}=-Fdt^{2}+G(d\rho ^{2}+\rho ^{2}d\Omega ^{2}),  \label{22}
\end{equation}

with

\begin{equation}
F=\frac{1}{4\rho r_{0}}\left[ \frac{\left( \rho ^{2}-\rho _{h}^{2}\right)
^{2}}{\left( \rho +\rho _{h}\right) ^{2}+4r_{-}\rho }\right] ,  \label{23}
\end{equation}

\begin{equation}
G=\frac{r_{0}}{4\rho ^{3}}\left[ 4\rho r_{+}+(\rho -\rho _{h})^{2}\right] ,
\label{24}
\end{equation}

Once we rewrite the metric (22) as follows

\begin{equation}
ds^{2}=F(-dt^{2}+\Delta ),  \label{25}
\end{equation}

we obtain the dBH\ in the form of the Fermat metric \cite{Perlick}

\begin{equation}
\Delta =n^{2}(d\rho ^{2}+\rho ^{2}d\Omega ^{2}),  \label{26}
\end{equation}

in which, as mentioned before, $n$ is called the index of refraction or the
refractive index. For the dBH medium, we compute it as

\begin{equation}
n=\frac{r_{0}\sqrt{[(\rho -\rho _{h})^{2}+4\rho r_{+}][(\rho +\rho
_{h})^{2}+4\rho r_{-}]}}{\rho (\rho ^{2}-\rho _{h}^{2})},  \label{27}
\end{equation}

When we use the HJ equation (8) on the ICS metric (22), we get

\begin{equation}
\frac{-1}{F}(\partial _{t}I)^{2}+\frac{1}{G}(\partial _{\rho }I)^{2}+\frac{1%
}{G\rho ^{2}}\left[ (\partial _{\theta }I)^{2}+\frac{1}{\sin ^{2}\theta }%
(\partial _{\varphi }I)^{2}\right] +m^{2}=0,  \label{28}
\end{equation}

It is possible to obtain the solution to the above equation in the following
form

\begin{equation}
I=-Et+W_{iso}(\rho )+J(x^{i}),  \label{29}
\end{equation}

Thus we get an integral solution for $W_{iso}(\rho )$\ as

\begin{equation}
W_{iso}(\rho )=\pm \int n\sqrt{E^{2}-\frac{F}{G\rho ^{2}}\left( J_{\theta
}^{2}+\frac{J_{\varphi }^{2}}{\sin ^{2}\theta }\right) -m^{2}F}d\rho ,
\label{30}
\end{equation}

Around the horizon the above solution reduces to

\begin{equation}
W_{iso}(\rho )\cong W_{iso(\pm )}=\pm E\int nd\rho ,  \label{31}
\end{equation}

One can easily observe that the $n$ of the dBH is decisive on the $%
W_{iso(\pm )}$. After making a straightforward calculation covering the
Feyman's prescription, we obtain

\begin{eqnarray}
W_{iso(\pm )} &=&\pm i2\pi Er_{0}\frac{\sqrt{r_{+}[\left( \rho
_{h}+r_{-}\right) ]}}{\rho _{h}}+\mu  \notag \\
&=&\pm i2\pi E\frac{r_{0}r_{+}}{r_{+}-r_{-}}+\mu ,  \label{32}
\end{eqnarray}

where $\mu $ is another complex integration constant. Similar to the
procedure followed in the previous section i.e., setting $P_{in}=1$ which
yields ${Im}\mu =2\pi Er_{0}\frac{r_{+}}{r_{+}-r_{-}}$, we derive the
imaginary part of the action $I$\ of the tunneling particle as follows

\begin{equation}
{Im}I={Im}W_{iso(+)}=4\pi E\frac{r_{0}r_{+}}{r_{+}-r_{-}},  \label{33}
\end{equation}

Thus, by considering the tunneling rate expresions (18) and (19) one obtains
the horizon temperature of the dBH as

\begin{equation}
\check{T}_{H}=\frac{r_{+}-r_{-}}{8\pi r_{+}r_{0}}.  \label{34}
\end{equation}

As it can be seen above, the result is the half of the standard Hawking
temperature; $\check{T}_{H}=\frac{1}{2}T_{H}$. Therefore, we deduce that the
present result infers that ICS leads to an apparent temperature of the dBH
that it is less than its $T_{H}$. This phenomenon resembles in the case that
the apparent depth $h$ of a fish swimming at a depth $d$ below the surface
of a pool is less than the true depth $d$. Namely, $h<d$. This illusion is
due to the discrepancy of the refractive indexes between the mediums. In
particular, such events take place when $n_{observer}<n_{object},$ as
happened at here. Because, it can be checked from Eq. (27) that the $n$
value of the medium of an observer who is placed at the outer region ($\rho
>\rho _{h}$) is less than the $n$ for the medium of the near the horizon ($%
\rho \approx \rho _{h}$). Since the value of $W_{iso(\pm )}$ (31) acts as a
decision-maker on the value of the horizon temperature $\check{T}_{H}$\ of
the dBH, one can deduce that the $n$ (27), and consequently the
gravitational lensing effect, plays an important role on the observation of
the true $T_{H}$.

On the other hand, we admittedly know that coordinate transformation of the
naive coordinates to the isotropic coordinates should not alter the true
temperature of the BH. Since the appearances are deceptive, one should make
deeper analysis to get the real. This problem has been recently discussed by
Chatterjee and Mitra \cite{GRG12} for the Schwarzschild BH and by Sakalli
and Mirekhtiary \cite{SM13} for the LDBH. They have proven that during the
evaluation of the integral (31) around the horizon, the path across the
horizon involves a change of $\pi /2$ instead of $\pi $. By getting inspired
from their studies, here we shall follow the same strategy to overcome that
factor-2 problem. For this purpose, we rewrite Eq. (21) as follows

\begin{eqnarray}
r &=&r_{-}+\frac{1}{4\rho }\left( \rho +\rho _{h}\right) ^{2},  \notag \\
&=&r_{+}+\frac{1}{4\rho }\left( \rho -\rho _{h}\right) ^{2},  \label{35}
\end{eqnarray}

which means that

\begin{equation}
\frac{dr}{r-r_{+}}=-\frac{d\rho }{\rho }+\frac{2d\rho }{\rho -\rho _{h}}.
\label{36}
\end{equation}

The first term at the right hand side (RHS) of the above equation does not
admit any imaginary part at the horizon. Only the second term of the RHS of
the Eq. (36) admits the imaginary result. On the other hand, in order to
match the RHS and the left hand side results any imaginary contribution
coming from $\frac{d\rho }{\rho -\rho _{h}}$ must be half of the $\frac{dr}{%
r-r_{+}}$. The latter remark produces a factor $i\pi /2$ for the integral
(31) and therefore it results in ${Im}W_{iso(+)}=2\pi E\frac{r_{0}r_{+}}{%
r_{+}-r_{-}}$ as obtained in the previous section. Thus, we read the horizon
temperature as $\check{T}_{H}=\frac{r_{+}-r_{-}}{4\pi r_{+}r_{0}}$ which is
nothing but the $T_{H}$.

\section{Conclusion}

In this study, the HR of the dBH which has double horizons is studied via
the HJ method. We have first applied the associated HJ\ method to the dBH\
metric expressed with the naive coordinates. Thus, we have explicitly shown
how the HJ\ method produces the $T_{H}$. On the other hand, the most
interesting part of the present paper belongs to the Sec. III in which the
dBH spacetime is transformed into the ICS. In this system, with the aid of
the Fermat metric we have managed to determine the $n$ of the dBH. In
particular, it is proven that the $n$ plays a decisive role on the tunneling
rate. Unlike to the naive coordinates, in the ICS the integration around the
pole which appears at the horizon has led the factor-2 problem in the
horizon temperature. For fixing this discrepancy, we inspired from recent
studies \cite{GRG12} \ and \cite{SM13} which have demonstrated how the
proper regularization of singular integrals give the true horizon
temperature, i.e., $T_{H}$, in the ICS. As a result, it has been clarified
that the path across the horizon entails the value of $\frac{i\pi }{2}$ on
the integration instead of $i\pi $. Our results are in accordance with the
work of \cite{SM13} if one gets the limit $r_{-}\rightarrow 0$.

\end{document}